\documentclass[twocolumn]{aastex631}
\usepackage[utf8]{inputenc}
\usepackage{natbib}
\usepackage{graphicx}
\usepackage{xcolor}
\usepackage{threeparttable}
\usepackage{multirow}

\received{}
\revised{}
\accepted{}
\submitjournal{Acta Prima Aprilia}

\shorttitle{New Constraints on the Cosmic Shoreline}
\shortauthors{Radica et al.}

\begin{document}

\title{New Constraints on the M Dwarf Cosmic Shoreline from a Galaxy Far, Far Away}

\correspondingauthor{Michael Radica}
\email{radicamc@uchicago.edu}

\author[0000-0002-3328-1203]{Michael Radica}
\affiliation{Department of Astronomy \& Astrophysics, University of Chicago, 5640 South Ellis Avenue, Chicago, IL 60637, USA}

\begin{abstract}
Whether there is a cosmic shoreline that divides terrestrial planets which have atmospheres from those that don't is one of the biggest open questions in exoplanet science. Most atmosphere searches have focused on terrestrial planets around M dwarf stars, since their smaller radii compared to sun-like stars boost planet atmosphere signals. However, the higher activity levels of M dwarfs might also entirely preclude atmosphere retention for their planets. In this work we present a new hope for defining an M dwarf cosmic shoreline, leveraging not only data from exoplanets in our own galaxy, but a comprehensive survey conducted by a commission of the Galactic Republic a long, long time ago in a galaxy far, far away. In this survey, we find definitive proof that M dwarf planets can retain atmospheres, and define an M dwarf cosmic shoreline whose slope agrees well with empirical predictions for Sun-like stars. We then define atmosphere retention metrics for the planets on the JWST Rocky Worlds DDT Targets Under Consideration list. Our analysis highlights the benefits of looking beyond the Milky Way for answers to some of the field's most pressing questions.
\end{abstract}

\keywords{Exoplanets (498); Exoplanet atmospheres (487); Planetary atmospheres (1244)}

\section{Introduction} 
\label{sec: Introduction}

The quest to locate an atmosphere on a small, potentially-Earth like planet outside of the solar system is one of the defining pursuits of exoplanet astronomy. To date, the majority of terrestrial atmosphere searches have focused on planets orbiting M dwarfs stars since their smaller sizes and lower luminosities compared to Sun-like stars amplify the planet-to-star radius ratio, and move their habitable zones to closer orbital distances such as can be probed with transit or eclipse observations \citep{trappist-1_jwst_community_initiative_roadmap_2024}. However, M dwarf stars are also significantly more active than Sun-like stars and possess longer pre-main sequence phases \citep{Wheatley_strong_2017}. It is thus unclear whether, if they did originally form with an atmosphere, a terrestrial M dwarf planet can retain said atmosphere to the present day.

To date, the vast majority of atmosphere searches, particularly for M dwarf terrestrials have been negative or inconclusive (but leaning towards negative). Emission studies with the Spitzer Space Telescope, or JWST's MIRI instrument have found bare rocks across the board \citep[e.g.,][]{kreidberg_absence_2019, greene_thermal_2023, zieba_no_2023, weiner_mansfield_no_2024, zhang_gj_2024, xue_jwst_2025}. Transit observations have generally been negative or inconclusive \citep[e.g.,][]{lim_atmospheric_2023, moran_high_2023, radica_promise_2025, wachiraphan_thremal_2025, allen_JWST_2025}, largely due to the impacts of stellar contamination in the form of the transit light source effect \citep[TLS; e.g.,][]{rackham_transit_2019}. However, there have been some tentative atmosphere signals identified which require further study \citep[e.g.,][]{damiano_lhs_2024, cadieux_transmission_2024, gressier_hints_2024, bello-arufe_evidence_2025}.

In all, the above factors have prevented the field from defining a so-called ``cosmic shoreline''. The concept of a cosmic shoreline was introduced by \citet{zahnle_cosmic_2017} as a line dividing objects which do possess atmospheres from those that don't in escape velocity-instellation space. Though the \citet{zahnle_cosmic_2017} shoreline was derived with reference to solar system bodies, the question naturally arises of whether this cosmic shoreline is a universal phenomenon that applies more broadly --- to solar system bodies and exoplanets alike --- and whether it is consistent across stellar types, i.e., does an M dwarf cosmic shoreline exist and is it the same as a shoreline for Sun-like stars? Searching for atmospheres around terrestrial M dwarf planets attempts to simultaneously answer these questions.  

Several attempts have been made to empirically derive cosmic shorelines in the exoplanet population \citep[e.g.,][]{meni-gallardo_empirical_2025, berta_3D_2025, pass_receding_2025}. Additional benefits of such analyses are that they provide a framework to identify targets which are the most promising candidates to host atmospheres --- purely by nature of being located the deepest into the ``atmosphere-favoured'' side of the parameter space (i.e., having lower instellation and higher escape velocity). The lack of terrestrial exoplanets around M dwarfs (or any star for that matter) with firm atmosphere detections, though, greatly complicates this endeavour. 

So far, all attempts to define a cosmic shoreline have focused on planets within our own galaxy. Here, we posit that if one is to understand the cosmic shoreline one must study all it’s aspects, not just the dogmatic, narrow view of the Milky Way. In this work, we perform a reanalysis of the M dwarf cosmic shoreline, also incorporating a wealth of data collected a long, long time ago in a galaxy far, far away. The structure of this article is as follows: in Section~\ref{sec: Data Sourcing} we describe our data sourcing process, and in Section~\ref{sec: Analysis} our cosmic shoreline analysis. In Section~\ref{sec: Discussion} we discuss our findings and we conclude in Section~\ref{sec: conclusions}.

\section{Data Sourcing} 
\label{sec: Data Sourcing}

For this work, we source our planetary data directly from the primary publications wherever possible. For Milky Way exoplanets, we used the Exoplanet Archive \citep{christiansen_nasa_2025} for primary planet properties (e.g., mass, radius, orbital period). We use a radius cutoff of $R<2\,R_{\oplus}$, similar to that of \citet{pass_receding_2025}, to limit our analysis to potentially terrestrial planets. We also apply a cut in stellar type to only retain systems with an M dwarf host star. A literature search was then conducted to assess whether individual planets in the sample have had evidence for the presence (0/50) or lack (11/50) of an atmosphere. For targets like LHS 1140\,b, TRAPPIST-1\,d and e which are inconclusive \citep[e.g.,][]{damiano_lhs_2024, allen_JWST_2025, piaulet_strict_2025}, we have not categorized them either way. 

\begin{figure}
    \centering
    \includegraphics[width=0.9\linewidth]{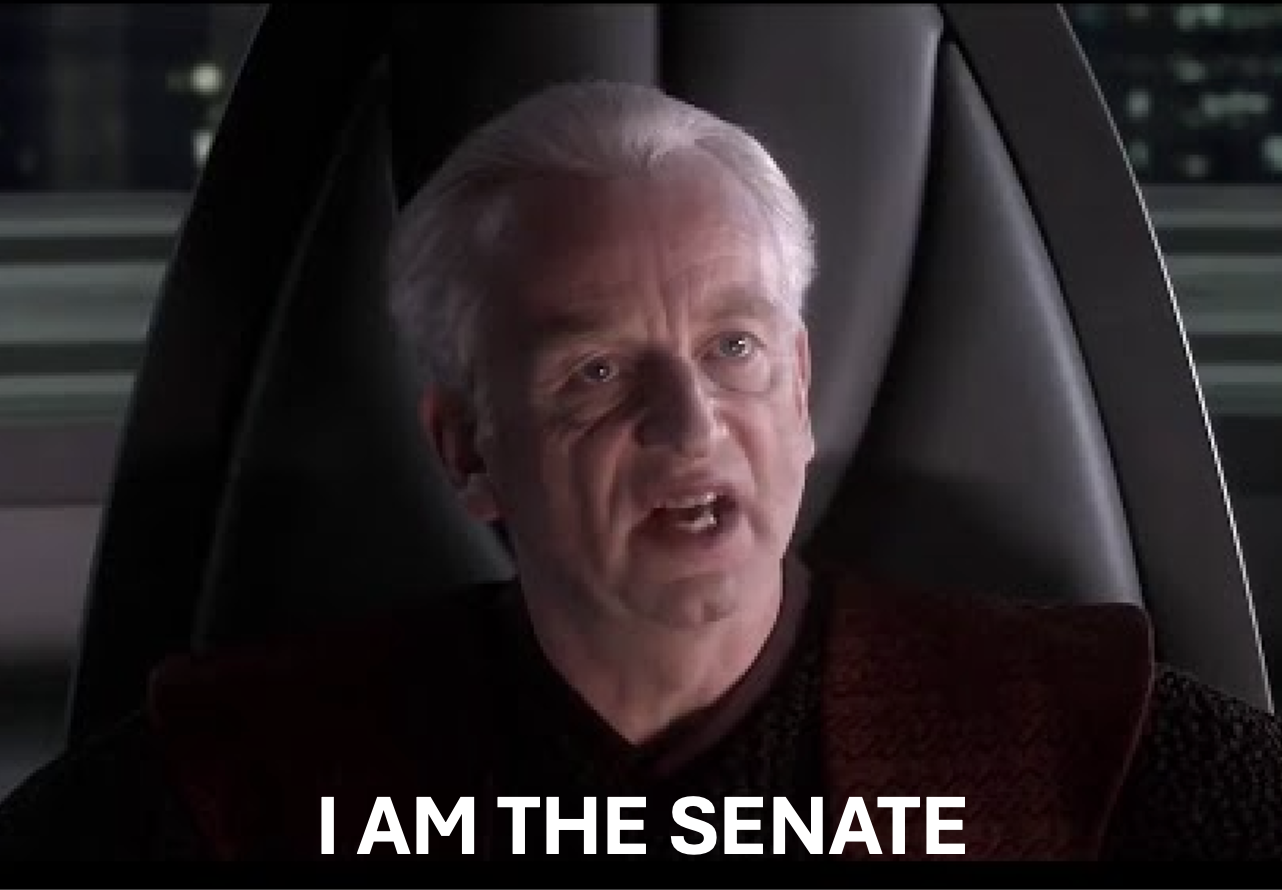}
    \caption{The Galactic Senate, circa 19 BBY, colourized.}
    \label{fig:senate}
\end{figure}

For extragalactic data, we make use of the great survey carrier out by the Galactic Senate (Figure~\ref{fig:senate}) during the administration of Supreme Chancellor Valorum in 41 BBY (Before the Battle of Yavin). Though we call this sample the ``extragalactic sample'' for simplicity, we note that it is sourced entirely from only one single other galaxy located far, far away from ours. We sourced the relevant planetary info from the Wookiepedia database\footnote{\url{https://starwars.fandom.com/wiki/Main_Page}}, applying the same physical cuts as for the Milky Way sample. This yielded a total of eight planets to add to the analysis. We also note for ethics purposes that no Bothans were harmed in the collection of this data. 

\begin{figure*}
    \centering
    \includegraphics[width=0.85\linewidth]{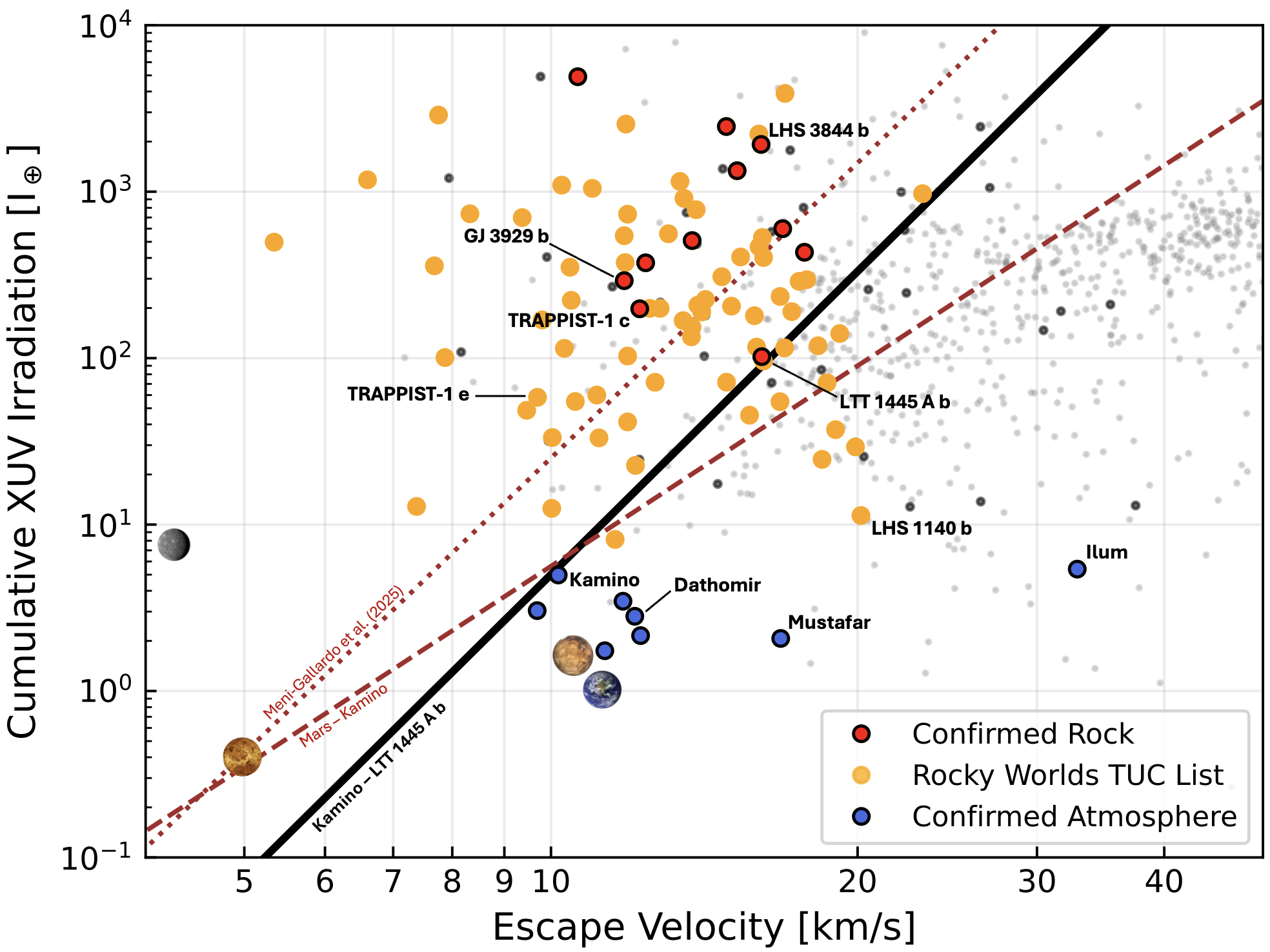}
    \caption{The M dwarf cosmic shoreline in XUV irradiation and escape velocity space. Small grey points show the broader exoplanet population, whereas darker points are potentially terrestrial planets around M dwarfs. Larger orange circles denote the Rocky Worlds DDT targets under consideration (TUC) list. Red points are bona fide rocks, and blue points have atmospheres. Several key planets are labelled, both from our galaxy and another far, far away. We show three potential empirical shorelines: one anchored by Mars and Kamino (red dashed), another anchored by LTT 1445\,A\,b and Kamino (black), and the shoreline for Sun-like stars derived by \citet{meni-gallardo_empirical_2025} (red dotted). }
    \label{fig:shoreline}
\end{figure*}

The entirety of our sample is visualized in Figure~\ref{fig:shoreline} in XUV irradiation and escape velocity space. Planets with atmosphere detections are marked in blue and bona fide rocks are in red. The cumulative XUV irradiation values are calculated following \citet{zahnle_cosmic_2017}, though future analyses should consider the updates from \citet{pass_receding_2025} for mid-to-late M dwarfs.

\section{Shoreline Analysis} 
\label{sec: Analysis}

Since the extragalactic sample provides the first firm evidence for terrestrial planets around M dwarfs retaining atmospheres, we can now derive a true M dwarf cosmic shoreline and compare it with shoreline derivations for Sun-like stars.

To this end, we use the linear support vector machine framework implemented in \texttt{scikit-learn} to define the line dividing the M dwarf atmosphere and atmosphere-less populations in XUV irradiation-escape velocity space. The two most constraining planets in this regard are LTT 1445\,A\,b on the atmosphere-less side and Kamino on the atmosphere side. We derive a shoreline which follows the equation
\begin{equation}
\label{equ: K-L}
    \log_{10} I_{XUV} = 6.04\log_{10} v_{esc} - 5.35
\end{equation}
and is show in black in Figure~\ref{fig:shoreline}. Hereafter, we refer to this as the Kamino-LTT 1445\,A\,b, or L-K, shoreline.

We then derive a second shoreline, but this time including the solar system planets. This has the effect of mixing stellar populations (i.e., the Sun with M dwarfs). Here we derive a shoreline of
\begin{equation}
\label{equ: M-K}
    \log_{10} I_{XUV} = 4.02\log_{10} v_{esc} - 3.21
\end{equation}
shown in a dashed red line in Figure~\ref{fig:shoreline}. Hereafter, we refer to this as the Mars-Kamino, or M-K, shoreline.

\section{Discussion} 
\label{sec: Discussion}

We find that our K-L shoreline has a comparable slope, but is offset in XUV Irradiation from the shoreline derived by \citet{meni-gallardo_empirical_2025} for mainly Sun-like stars using Mars and 55 Cancri e as anchors (e.g., Figure~\ref{fig:shoreline}). They obtain a slope of 5.89 and an intercept of $-4.49$. This potentially indicates that it is indeed the same physical process that dictates atmosphere retention around M dwarf and Sun-like stars (on account of the similarity in slope) but that atmosphere removal is more efficient around M dwarfs (since the y-intercept is at lower irradiation values). This is entirely consistent with M dwarfs having higher levels of XUV radiation compared to Sun-like stars \citep{Wheatley_strong_2017} and spending longer periods of their lifetimes outputting this high-energy radiation that primarily drives atmosphere loss \citep{pass_receding_2025}.

\begin{deluxetable*}{c|cccccccc}
 \centering
 \tabletypesize{\footnotesize}
 \label{tab: ARMs}
 \tablecaption{Atmosphere Retention Metrics and System Properties for M Dwarf Terrestrial Planets from the JWST Rocky Worlds TUC List}
 \tablehead{
 Planet & $M_*$ & $L\rm _{bol}$ & $a$ & $v \rm_{esc}$ & $I\rm _{XUV}$ & ARM & ARM & Mass?\\
  & [$M_{\odot}$] & [$\log L_{\odot}$] & [$au$] & [$km/s$] & [$I_{\oplus}$] & L-K & M-K &
  }
    \startdata
     TOI-2285\,b & 0.45 & -1.55 & 0.136 & 37.49 & 13.02 & 3.06 & 1.93 & N \\
     TOI-2095\,c & 0.44 & -1.46 & 0.137 & 26.42 & 13.75 & 2.11 & 1.30 & N \\
     LHS 1140\,b & 0.18 & -2.37 & 0.094 & 22.52 & 12.79 & 1.73 & 1.05 & Y \\
     LP 890-9\,b & 0.12 & -2.82 & 0.019 & 35.42 & 210.01 & 1.70 & 0.62 & N \\
     K2-415\,b & 0.16 & -2.44 & 0.027 & 30.44 & 146.98 & 1.46 & 0.52 & N \\
     LHS 1815\,b & 0.5 & -1.40 & 0.040 & 31.67 & 191.34 & 1.45 & 0.47 & N \\
     TOI-2095\,b & 0.44 & -1.46 & 0.101 & 20.28 & 25.62 & 1.15 & 0.57 & N \\
     K2-3\,c & 0.55 & -1.23 & 0.136 & 14.58 & 17.55 & 0.45 & 0.16 & Y \\
     TOI-1235\,b & 0.64 & -1.09 & 0.038 & 22.33 & 246.61 & 0.42 & -0.25 & Y \\
     TOI-663\,d & 0.51 & -1.37 & 0.058 & 18.43 & 85.14 & 0.38 & -0.12 & N \\
     TOI-1634\,b & 0.45 & -1.58 & 0.015 & 26.97 & 1056.16 & 0.28 & -0.55 & Y \\
     TOI-1695\,b & 0.51 & -1.35 & 0.034 & 20.49 & 258.41 & 0.17 & -0.42 & Y \\
     TOI-776\,b & 0.54 & -1.30 & 0.065 & 16.47 & 71.16 & 0.16 & -0.24 & Y \\
     TOI-5720\,b & 0.38 & -1.80 & 0.018 & 22.25 & 587.16 & 0.03 & -0.63 & N \\
     TOI-244\,b & 0.43 & -1.64 & 0.056 & 14.87 & 70.52 & -0.11 & -0.41 & Y \\
     TOI-1075\,b & 0.6 & -1.17 & 0.012 & 26.40 & 2454.49 & -0.14 & -0.95 & Y \\
     HD 260655\,c & 0.44 & -1.44 & 0.047 & 15.90 & 117.78 & -0.15 & -0.52 & Y \\
     TRAPPIST-1\,g & 0.08 & -3.25 & 0.045 & 12.21 & 24.59 & -0.17 & -0.29 & Y \\
     TOI-6008\,b & 0.23 & -2.33 & 0.011 & 22.07 & 995.72 & -0.22 & -0.87 & N \\
     LTT 1445\,A\,b & 0.26 & -2.06 & 0.038 & 14.14 & 102.77 & -0.40 & -0.66 & Y \\
     L 98-59\,d & 0.27 & -1.96 & 0.049 & 12.65 & 70.81 & -0.53 & -0.69 & Y \\
     TRAPPIST-1\,f & 0.09 & -3.25 & 0.038 & 11.17 & 33.78 & -0.54 & -0.59 & Y \\
     L 98-59\,c & 0.27 & -1.96 & 0.030 & 14.18 & 180.68 & -0.64 & -0.90 & Y \\
     LHS 1140\,c & 0.18 & -2.40 & 0.027 & 13.72 & 153.06 & -0.65 & -0.88 & Y \\
     GJ 357\,b & 0.34 & -1.80 & 0.035 & 13.77 & 164.15 & -0.67 & -0.91 & Y \\
     L 168-9\,b & 0.61 & -1.15 & 0.021 & 17.70 & 802.22 & -0.70 & -1.16 & Y \\
     TOI-406\,c & 0.41 & -1.70 & 0.032 & 14.06 & 201.04 & -0.71 & -0.96 & Y \\
     GJ 486\,b & 0.32 & -1.91 & 0.017 & 16.46 & 574.89 & -0.75 & -1.14 & Y \\
     LHS 1478\,b & 0.24 & -2.13 & 0.018 & 15.34 & 389.29 & -0.77 & -1.10 & Y \\
     LHS 1678\,c & 0.34 & -1.84 & 0.033 & 13.37 & 172.91 & -0.77 & -0.98 & N \\
     HD 260655\,b & 0.44 & -1.44 & 0.029 & 14.71 & 308.76 & -0.77 & -1.07 & Y \\
     GJ 12\,b & 0.24 & -2.13 & 0.067 & 9.95 & 31.79 & -0.82 & -0.76 & Y \\
     TOI-270\,b & 0.39 & -1.71 & 0.032 & 12.82 & 201.92 & -0.95 & -1.12 & Y \\
     TRAPPIST-1\,c & 0.08 & -3.25 & 0.015 & 12.80 & 215.86 & -0.98 & -1.15 & Y \\
     GJ 1132\,b & 0.19 & -2.32 & 0.016 & 13.90 & 484.82 & -1.12 & -1.36 & Y \\
     TOI-1685\,b & 0.46 & -1.56 & 0.012 & 17.17 & 1775.15 & -1.13 & -1.56 & Y \\
     TRAPPIST-1\,e & 0.09 & -3.25 & 0.029 & 9.71 & 58.47 & -1.14 & -1.07 & Y \\
     TRAPPIST-1\,h & 0.09 & -3.25 & 0.062 & 7.36 & 13.07 & -1.22 & -0.90 & Y \\
     GJ 3929\,b & 0.31 & -1.92 & 0.026 & 11.49 & 268.62 & -1.36 & -1.44 & Y \\
     GJ 3473\,b & 0.36 & -1.83 & 0.016 & 13.59 & 749.06 & -1.37 & -1.59 & Y \\
     Wolf 327\,b & 0.41 & -1.62 & 0.010 & 16.00 & 2228.36 & -1.41 & -1.78 & Y \\
     LTT 3780\,b & 0.38 & -1.78 & 0.012 & 14.75 & 1372.52 & -1.42 & -1.71 & Y \\
     LP 791-18\,d & 0.14 & -2.65 & 0.020 & 10.46 & 218.71 & -1.52 & -1.51 & Y \\
     GJ 806\,b & 0.41 & -1.59 & 0.014 & 13.38 & 1206.72 & -1.62 & -1.83 & Y \\
     GJ 1252\,b & 0.38 & -1.71 & 0.009 & 14.82 & 2480.75 & -1.66 & -1.96 & Y \\
     TRAPPIST-1\,d & 0.08 & -3.25 & 0.021 & 8.16 & 108.76 & -1.87 & -1.64 & Y \\
     TRAPPIST-1\,b & 0.08 & -3.25 & 0.011 & 9.91 & 404.95 & -1.93 & -1.87 & Y \\
     L 98-59\,b & 0.27 & -1.96 & 0.022 & 7.68 & 348.88 & -2.54 & -2.25 & Y \\
     LHS 1678\,b & 0.34 & -1.84 & 0.013 & 7.94 & 1206.96 & -2.99 & -2.73 & N \\
     GJ 367\,b & 0.45 & -1.53 & 0.007 & 9.77 & 4913.12 & -3.05 & -2.98 & Y \\
    \enddata
    \tablecomments{L-K denotes the cosmic shoreline defined using LTT 1445\,A\,b and Kamino as anchors whereas M-K uses Mars and Kamino. A radius cutoff of $R<2\,R_{\oplus}$ is used to limit the sample to potentially terrestrial planets, following \citet{pass_receding_2025}. A ``N'' in the Mass? column indicates the planet does not yet have a measured mass.}
\end{deluxetable*}

Compared to the M-K shoreline, however, we see that both the slope and intercept are very different from the K-L M dwarf shoreline derived here or the Sun-like star shoreline from \citet{meni-gallardo_empirical_2025}. This is likely due to the mixing of stellar types inherent in the M-K analysis, and potential indicates that cosmic shoreline analyses should only be conducted on a single stellar type. Alternatively, a new, normalized Y-axis which also takes into account stellar type (i.e., normalizing by XUV fraction or saturated XUV lifetime; \citealp{pass_receding_2025}) may be a fruitful avenue to combine different stellar types, though that is beyond the scope of the current work. 

We then proceed to follow several recent works \citep[e.g.,][]{pass_receding_2025, meni-gallardo_empirical_2025} and define an atmosphere retention metric (ARM) based on the shorelines derived in this work. We derive one ARM value for the K-L and one for the M-L shoreline following
\begin{equation}
\label{equ: ARM}
    ARM_{K-L} = 6.04\log_{10} v_{esc} - \log_{10} I_{XUV} + 5.35
\end{equation}
For the M-L shoreline, the slope and y-intercept from equation~\ref{equ: M-K} are used instead of those from equation~\ref{equ: K-L}. Larger positive values indicate a strong possibility for atmosphere retention, whereas large negative values indicate a high likelihood of the planet being a bare rock. The results for the planets in the targets under consideration (TUC) list from the JWST Rocky Worlds DDT program, as well as relevant system parameters sourced from the Exoplanet Archive, are shown in Table~\ref{tab: ARMs}. In cases where planets did not have a firm mass measurement (denoted with an ``N'' in the Mass? column), Exoplanet Archive upper limits were used for the escape velocity calculation. These calculations also do not take into account system ``individuality'' (an unusually low XUV output \citep[e.g.,][]{taylor_JWST_2025}, a planet-killing super-weapon actively draining the star, etc.).

Like several previous analyses we find that LHS 1140\,b is a very promising target for atmosphere retention based on its ARM value. Future JWST observations may potentially be able to confirm the tentative atmosphere signatures from \citet{damiano_lhs_2024} and \citet{cadieux_transmission_2024}. Our analysis also identifies multiple other Milky Way planets with very promising ARM values --- some of which are even larger than that of LHS 1140\,b. However, most of these do not currently have published mass measurements, either from radial velocity or transit timing variation analyses. It is thus possible that their ARM values are biased high through the use of mass upper limits as if they were true planet masses. However, this still indicates that these planets have the potential to be promising candidates for atmosphere retention and should be prioritized for followup.

\section{Conclusions}
\label{sec: conclusions}

In this paper we attempted the first-ever cosmic shoreline analysis that combines observations both from our galaxy and another far, far away. We derive an M dwarf shoreline from the combined Milky Way and extragalactic sample that shows a consistent slope, but smaller y-intercept to shorelines calculated for the Milky Way Sun-like population --- suggesting that the processes that drive atmosphere loss or retention are shared across stellar types, but loss processes are perhaps more efficient for later-type stars. We then calculated ARM values, encapsulating the likelihood of atmosphere retention, for the JWST Rocky Worlds DDT TUC list based on our derived cosmic shoreline. This analysis highlights LHS 1140\,b, as well as several other planets without firm mass measurements as promising targets in the quest to identify atmospheres around terrestrial M dwarf planets within our galaxy. 

In all, our work brings a new hope to the quest to derive an M dwarf cosmic shoreline. We encourage future works to reanalyze our data to ensure that our derived shoreline is not merely a phantom menace, and perhaps incorporate data from surveys not considered here, even ones that have boldly gone where no [wo]man has gone before.


\vspace{5mm}
\facilities{Exoplanet Archive \citep{christiansen_nasa_2025}}

\software{\texttt{ipython} \citep{PER-GRA:2007},
\texttt{matplotlib} \citep{Hunter:2007},
\texttt{numpy} \citep{harris2020array},
\texttt{scikit-learn} \citep{scikit-learn}
}

\bibliography{main}{}

@Article{Hunter:2007,
    Author = {Hunter, J. D.},
    Title = {Matplotlib: A 2D graphics environment},
    Journal = {Computing in Science \& Engineering},
    Volume = {9},
    Number = {3},
    Pages = {90--95},
    publisher = {IEEE COMPUTER SOC},
    doi = {10.1109/MCSE.2007.55},
    year = 2007
}

@Article{PER-GRA:2007,
    Author = {P\'erez, Fernando and Granger, Brian E.},
    Title = {{IP}ython: a System for Interactive Scientific Computing},
    Journal = {Computing in Science and Engineering},
    Volume = {9},
    Number = {3},
    Pages = {21--29},
    month = may,
    year = 2007,
    url = "https://ipython.org",
    ISSN = "1521-9615",
    doi = {10.1109/MCSE.2007.53},
    publisher = {IEEE Computer Society},
}

@Article{harris2020array,
    title = {Array programming with {NumPy}},
    author = {Charles R. Harris and K. Jarrod Millman and St{\'{e}}fan J. van der Walt and Ralf Gommers and Pauli Virtanen and David Cournapeau and Eric Wieser and Julian Taylor and Sebastian Berg and Nathaniel J. Smith and Robert Kern and Matti Picus and Stephan Hoyer and Marten H. van Kerkwijk and Matthew Brett and Allan Haldane and Jaime Fern{\'{a}}ndez del R{\'{i}}o and Mark Wiebe and Pearu Peterson and Pierre G{\'{e}}rard-Marchant and Kevin Sheppard and Tyler Reddy and Warren Weckesser and Hameer Abbasi and Christoph Gohlke and Travis E. Oliphant},
    year = {2020},
    month = sep,
    journal = {Nature},
    volume = {585},
    number = {7825},
    pages = {357--362},
    doi = {10.1038/s41586-020-2649-2},
    publisher = {Springer Science and Business Media {LLC}},
    url = {https://doi.org/10.1038/s41586-020-2649-2}
}

@article{scikit-learn,
  title={Scikit-learn: Machine Learning in {P}ython},
  author={Pedregosa, F. and Varoquaux, G. and Gramfort, A. and Michel, V.
          and Thirion, B. and Grisel, O. and Blondel, M. and Prettenhofer, P.
          and Weiss, R. and Dubourg, V. and Vanderplas, J. and Passos, A. and
          Cournapeau, D. and Brucher, M. and Perrot, M. and Duchesnay, E.},
  journal={Journal of Machine Learning Research},
  volume={12},
  pages={2825--2830},
  year={2011}
}

@ARTICLE{Wheatley_strong_2017,
       author = {{Wheatley}, Peter J. and {Louden}, Tom and {Bourrier}, Vincent and {Ehrenreich}, David and {Gillon}, Micha{\"e}l},
        title = "{Strong XUV irradiation of the Earth-sized exoplanets orbiting the ultracool dwarf TRAPPIST-1}",
      journal = {Monthly Notices of the Royal Astronomical Society},
         year = 2017,
        month = feb,
       volume = {465},
       number = {1},
        pages = {L74-L78},
          doi = {10.1093/mnrasl/slw192},
archivePrefix = {arXiv},
       eprint = {1605.01564},
 primaryClass = {astro-ph.EP},
       adsurl = {https://ui.adsabs.harvard.edu/abs/2017MNRAS.465L..74W}
}

@ARTICLE{xue_jwst_2025,
       author = {{Xue}, Qiao and {Zhang}, Michael and {Coy}, Brandon Park and {Brady}, Madison and {Ji}, Xuan and {Bean}, Jacob L. and {Radica}, Michael and {Seifahrt}, Andreas and {St{\"u}rmer}, Julian and {Luque}, Rafael and {Basant}, Ritvik and {Brown}, Nina and {Das}, Tanya and {Kasper}, David and {Piaulet-Ghorayeb}, Caroline and {Kempton}, Eliza M.-R. and {Kite}, Edwin},
        title = "{The JWST Rocky Worlds DDT Program Reveals GJ 3929b to Likely Be a Bare Rock}",
      journal = {The Astrophysical Journal Letters},
         year = 2025,
        month = dec,
       volume = {995},
       number = {2},
          eid = {L52},
        pages = {L52},
          doi = {10.3847/2041-8213/ae2098},
archivePrefix = {arXiv},
       eprint = {2508.12516},
 primaryClass = {astro-ph.EP},
       adsurl = {https://ui.adsabs.harvard.edu/abs/2025ApJ...995L..52X}
}

@article{damiano_lhs_2024,
   title={LHS 1140 b Is a Potentially Habitable Water World},
   volume={968},
   ISSN={2041-8213},
   url={http://dx.doi.org/10.3847/2041-8213/ad5204},
   DOI={10.3847/2041-8213/ad5204},
   number={2},
   journal={The Astrophysical Journal Letters},
   publisher={American Astronomical Society},
   author={Damiano, Mario and Bello-Arufe, Aaron and Yang, Jeehyun and Hu, Renyu},
   year={2024},
   month=jun, pages={L22} 
}

@ARTICLE{allen_JWST_2025,
       author = {{Allen}, Natalie H. and {Espinoza}, N{\'e}stor and {Boehm}, V.~A. and {Ca{\~n}as}, Caleb I. and {Stevenson}, Kevin B. and {Lewis}, Nikole K. and {MacDonald}, Ryan J. and {Morris}, Brett M. and {Agol}, Eric and {Col{\'o}n}, Knicole and {Diamond-Lowe}, Hannah and {Glidden}, Ana and {Gressier}, Am{\'e}lie and {Huang}, Jingcheng and {Lin}, Zifan and {Long}, Douglas and {Louie}, Dana R. and {MacGregor}, Meredith A. and {Pueyo}, Laurent and {Rackham}, Benjamin V. and {Ranjan}, Sukrit and {Seager}, Sara and {Tovar Mendoza}, Guadalupe and {Valenti}, Jeff A. and {Valentine}, Daniel and {van der Marel}, Roeland P. and {Wakeford}, Hannah R.},
        title = "{JWST TRAPPIST-1 e/b Program: Motivation and First Observations}",
      journal = {The Astronomical Journal},
         year = 2026,
        month = feb,
       volume = {171},
       number = {2},
          eid = {105},
        pages = {105},
          doi = {10.3847/1538-3881/ae28cb},
archivePrefix = {arXiv},
       eprint = {2512.07695},
 primaryClass = {astro-ph.EP},
       adsurl = {https://ui.adsabs.harvard.edu/abs/2026AJ....171..105A}
}

@article{gressier_hints_2024,
   title={Hints of a Sulfur-rich Atmosphere around the 1.6 R
               ⊕ Super-Earth L98-59 d from JWST NIRspec G395H Transmission Spectroscopy},
   volume={975},
   ISSN={2041-8213},
   url={http://dx.doi.org/10.3847/2041-8213/ad73d1},
   DOI={10.3847/2041-8213/ad73d1},
   number={1},
   journal={The Astrophysical Journal Letters},
   publisher={American Astronomical Society},
   author={Gressier, Amélie and Espinoza, Néstor and Allen, Natalie H. and Sing, David K. and Banerjee, Agnibha and Barstow, Joanna K. and Valenti, Jeff A. and Lewis, Nikole K. and Birkmann, Stephan M. and Challener, Ryan C. and Manjavacas, Elena and Alves de Oliveira, Catarina and Crouzet, Nicolas and Beck, Tracy. L},
   year={2024},
   month=oct, pages={L10} 
}

@misc{berta_3D_2025,
      title={The 3D Cosmic Shoreline for Nurturing Planetary Atmospheres}, 
      author={Zach K. Berta-Thompson and Patcharapol Wachiraphan and Catriona Murray},
      year={2025},
      eprint={2507.02136},
      archivePrefix={arXiv},
      primaryClass={astro-ph.EP},
      url={https://arxiv.org/abs/2507.02136}, 
}

@article{wachiraphan_thremal_2025,
   title={The Thermal Emission Spectrum of the Nearby Rocky Exoplanet LTT 1445A b from JWST MIRI/LRS},
   volume={169},
   ISSN={1538-3881},
   url={http://dx.doi.org/10.3847/1538-3881/adc990},
   DOI={10.3847/1538-3881/adc990},
   number={6},
   journal={The Astronomical Journal},
   publisher={American Astronomical Society},
   author={Wachiraphan, Patcharapol and Berta-Thompson, Zachory K. and Diamond-Lowe, Hannah and Winters, Jennifer G. and Murray, Catriona and Zhang, Michael and Xue, Qiao and Morley, Caroline V. and Rosario-Franco, Marialis and Duvvuri, Girish M.},
   year={2025},
   month=may, pages={311} 
}

@ARTICLE{piaulet_strict_2025,
       author = {{Piaulet-Ghorayeb}, Caroline and {Benneke}, Bj{\"o}rn and {Turbet}, Martin and {Moore}, Keavin and {Roy}, Pierre-Alexis and {Lim}, Olivia and {Doyon}, Ren{\'e} and {Fauchez}, Thomas J. and {Albert}, Lo{\"\i}c and {Radica}, Michael and {Coulombe}, Louis-Philippe and {Lafreni{\`e}re}, David and {Cowan}, Nicolas B. and {Belzile}, Danika and {Musfirat}, Kamrul and {Kaur}, Mehramat and {L'Heureux}, Alexandrine and {Johnstone}, Doug and {MacDonald}, Ryan J. and {Allart}, Romain and {Dang}, Lisa and {Kaltenegger}, Lisa and {Pelletier}, Stefan and {Rowe}, Jason F. and {Taylor}, Jake and {Turner}, Jake D.},
        title = "{Strict Limits on Potential Secondary Atmospheres on the Temperate Rocky Exo-Earth TRAPPIST-1 d}",
      journal = {The Astrophysical Journal},
         year = 2025,
        month = aug,
       volume = {989},
       number = {2},
          eid = {181},
        pages = {181},
          doi = {10.3847/1538-4357/adf207},
archivePrefix = {arXiv},
       eprint = {2508.08416},
 primaryClass = {astro-ph.EP},
       adsurl = {https://ui.adsabs.harvard.edu/abs/2025ApJ...989..181P}
}

@ARTICLE{taylor_JWST_2025,
       author = {{Taylor}, Jake and {Radica}, Michael and {Chatterjee}, Richard D. and {Hammond}, Mark and {Meier}, Tobias and {Aigrain}, Suzanne and {MacDonald}, Ryan J. and {Albert}, Loic and {Benneke}, Bj{\"o}rn and {Coulombe}, Louis-Philippe and {Cowan}, Nicolas B. and {Dang}, Lisa and {Doyon}, Ren{\'e} and {Flagg}, Laura and {Johnstone}, Doug and {Kaltenegger}, Lisa and {Lafreni{\`e}re}, David and {Pelletier}, Stefan and {Piaulet-Ghorayeb}, Caroline and {Rowe}, Jason F. and {Roy}, Pierre-Alexis},
        title = "{JWST NIRISS transmission spectroscopy of the super-Earth GJ 357b, a favourable target for atmospheric retention}",
      journal = {Monthly Notices of the Royal Astronomical Society},
         year = 2025,
        month = jul,
       volume = {540},
       number = {4},
        pages = {3677-3692},
          doi = {10.1093/mnras/staf894},
archivePrefix = {arXiv},
       eprint = {2505.24462},
 primaryClass = {astro-ph.EP},
       adsurl = {https://ui.adsabs.harvard.edu/abs/2025MNRAS.540.3677T}
}

@article{lim_atmospheric_2023,
	title = {Atmospheric {Reconnaissance} of {TRAPPIST}-1 b with {JWST}/{NIRISS}: {Evidence} for {Strong} {Stellar} {Contamination} in the {Transmission} {Spectra}},
	volume = {955},
	issn = {2041-8205, 2041-8213},
	shorttitle = {Atmospheric {Reconnaissance} of {TRAPPIST}-1 b with {JWST}/{NIRISS}},
	url = {https://iopscience.iop.org/article/10.3847/2041-8213/acf7c4},
	doi = {10.3847/2041-8213/acf7c4},
	language = {en},
	number = {1},
	urldate = {2023-10-02},
	journal = {ApJL},
	author = {Lim, Olivia and Benneke, Björn and Doyon, René and MacDonald, Ryan J. and Piaulet, Caroline and Artigau, Étienne and Coulombe, Louis-Philippe and Radica, Michael and L’Heureux, Alexandrine and Albert, Loïc and Rackham, Benjamin V. and De Wit, Julien and Salhi, Salma and Roy, Pierre-Alexis and Flagg, Laura and Fournier-Tondreau, Marylou and Taylor, Jake and Cook, Neil J. and Lafrenière, David and Cowan, Nicolas B. and Kaltenegger, Lisa and Rowe, Jason F. and Espinoza, Néstor and Dang, Lisa and Darveau-Bernier, Antoine},
	month = sep,
	year = {2023},
	pages = {L22},
}

@article{rackham_transit_2019,
	title = {The {Transit} {Light} {Source} {Effect}. {II}. {The} {Impact} of {Stellar} {Heterogeneity} on {Transmission} {Spectra} of {Planets} {Orbiting} {Broadly} {Sun}-like {Stars}},
	volume = {157},
	issn = {1538-3881},
	url = {https://iopscience.iop.org/article/10.3847/1538-3881/aaf892},
	doi = {10.3847/1538-3881/aaf892},
	language = {en},
	number = {3},
	urldate = {2023-10-19},
	journal = {AJ},
	author = {Rackham, Benjamin V. and Apai, Dániel and Giampapa, Mark S.},
	month = feb,
	year = {2019},
	pages = {96},
}

@article{greene_thermal_2023,
	title = {Thermal emission from the {Earth}-sized exoplanet {TRAPPIST}-1 b using {JWST}},
	volume = {618},
	issn = {0028-0836, 1476-4687},
	url = {https://www.nature.com/articles/s41586-023-05951-7},
	doi = {10.1038/s41586-023-05951-7},
	language = {en},
	number = {7963},
	urldate = {2024-04-29},
	journal = {Nature},
	author = {Greene, Thomas P. and Bell, Taylor J. and Ducrot, Elsa and Dyrek, Achrène and Lagage, Pierre-Olivier and Fortney, Jonathan J.},
	month = jun,
	year = {2023},
	pages = {39--42},
}

@article{kreidberg_absence_2019,
	title = {Absence of a thick atmosphere on the terrestrial exoplanet {LHS} 3844b},
	volume = {573},
	issn = {0028-0836, 1476-4687},
	url = {https://www.nature.com/articles/s41586-019-1497-4},
	doi = {10.1038/s41586-019-1497-4},
	language = {en},
	number = {7772},
	urldate = {2024-04-29},
	journal = {Nature},
	author = {Kreidberg, Laura and Koll, Daniel D. B. and Morley, Caroline and Hu, Renyu and Schaefer, Laura and Deming, Drake and Stevenson, Kevin B. and Dittmann, Jason and Vanderburg, Andrew and Berardo, David and Guo, Xueying and Stassun, Keivan and Crossfield, Ian and Charbonneau, David and Latham, David W. and Loeb, Abraham and Ricker, George and Seager, Sara and Vanderspek, Roland},
	month = sep,
	year = {2019},
	pages = {87--90},
}

@article{zieba_no_2023,
	title = {No thick carbon dioxide atmosphere on the rocky exoplanet {TRAPPIST}-1 c},
	volume = {620},
	issn = {0028-0836, 1476-4687},
	url = {https://www.nature.com/articles/s41586-023-06232-z},
	doi = {10.1038/s41586-023-06232-z},
	language = {en},
	number = {7975},
	urldate = {2024-04-29},
	journal = {Nature},
	author = {Zieba, Sebastian and Kreidberg, Laura and Ducrot, Elsa and Gillon, Michaël and Morley, Caroline and Schaefer, Laura and Tamburo, Patrick and Koll, Daniel D. B. and Lyu, Xintong and Acuña, Lorena and Agol, Eric and Iyer, Aishwarya R. and Hu, Renyu and Lincowski, Andrew P. and Meadows, Victoria S. and Selsis, Franck and Bolmont, Emeline and Mandell, Avi M. and Suissa, Gabrielle},
	month = aug,
	year = {2023},
	pages = {746--749},
}

@article{zahnle_cosmic_2017,
	title = {The {Cosmic} {Shoreline}: {The} {Evidence} that {Escape} {Determines} which {Planets} {Have} {Atmospheres}, and what this {May} {Mean} for {Proxima} {Centauri} {B}},
	volume = {843},
	issn = {0004-637X, 1538-4357},
	shorttitle = {The {Cosmic} {Shoreline}},
	url = {https://iopscience.iop.org/article/10.3847/1538-4357/aa7846},
	doi = {10.3847/1538-4357/aa7846},
	language = {en},
	number = {2},
	urldate = {2024-04-29},
	journal = {ApJ},
	author = {Zahnle, Kevin J. and Catling, David C.},
	month = jul,
	year = {2017},
	pages = {122},
}

@article{zhang_gj_2024,
	title = {{GJ} 367b {Is} a {Dark}, {Hot}, {Airless} {Sub}-{Earth}},
	volume = {961},
	issn = {2041-8205, 2041-8213},
	url = {https://iopscience.iop.org/article/10.3847/2041-8213/ad1a07},
	doi = {10.3847/2041-8213/ad1a07},
	language = {en},
	number = {2},
	urldate = {2024-04-29},
	journal = {ApJL},
	author = {Zhang, Michael and Hu, Renyu and Inglis, Julie and Dai, Fei and Bean, Jacob L. and Knutson, Heather A. and Lam, Kristine and Goffo, Elisa and Gandolfi, Davide},
	month = feb,
	year = {2024},
	pages = {L44},
}

@article{cadieux_transmission_2024,
	title = {Transmission {Spectroscopy} of the {Habitable} {Zone} {Exoplanet} {LHS} 1140 b with {JWST}/{NIRISS}},
	volume = {970},
	issn = {2041-8205, 2041-8213},
	url = {https://iopscience.iop.org/article/10.3847/2041-8213/ad5afa},
	doi = {10.3847/2041-8213/ad5afa},
	language = {en},
	number = {1},
	urldate = {2024-07-23},
	journal = {ApJL},
	author = {Cadieux, Charles and Doyon, René and MacDonald, Ryan J. and Turbet, Martin and Artigau, Étienne and Lim, Olivia and Radica, Michael and Fauchez, Thomas J. and Salhi, Salma and Dang, Lisa and Albert, Loïc and Coulombe, Louis-Philippe and Cowan, Nicolas B. and Lafrenière, David and L’Heureux, Alexandrine and Piaulet-Ghorayeb, Caroline and Benneke, Björn and Cloutier, Ryan and Charnay, Benjamin and Cook, Neil J. and Fournier-Tondreau, Marylou and Plotnykov, Mykhaylo and Valencia, Diana},
	month = jul,
	year = {2024},
	pages = {L2},
}

@article{trappist-1_jwst_community_initiative_roadmap_2024,
	title = {A roadmap for the atmospheric characterization of terrestrial exoplanets with {JWST}},
	volume = {8},
	issn = {2397-3366},
	url = {https://www.nature.com/articles/s41550-024-02298-5},
	doi = {10.1038/s41550-024-02298-5},
	language = {en},
	number = {7},
	urldate = {2024-08-15},
	journal = {Nat Astron},
	author = {{TRAPPIST-1 JWST Community Initiative} and De Wit, Julien and Doyon, René and Rackham, Benjamin V. and Lim, Olivia and Ducrot, Elsa and Kreidberg, Laura and Benneke, Björn and Ribas, Ignasi and Berardo, David and Niraula, Prajwal and Iyer, Aishwarya and Shapiro, Alexander and Kostogryz, Nadiia and Witzke, Veronika and Gillon, Michaël and Agol, Eric and Meadows, Victoria and Burgasser, Adam J. and Owen, James E. and Fortney, Jonathan J. and Selsis, Franck and Bello-Arufe, Aaron and De Beurs, Zoë and Bolmont, Emeline and Cowan, Nicolas and Dong, Chuanfei and Drake, Jeremy J. and Garcia, Lionel and Greene, Thomas and Haworth, Thomas and Hu, Renyu and Kane, Stephen R. and Kervella, Pierre and Koll, Daniel and Krissansen-Totton, Joshua and Lagage, Pierre-Olivier and Lichtenberg, Tim and Lustig-Yaeger, Jacob and Lingam, Manasvi and Turbet, Martin and Seager, Sara and Barkaoui, Khalid and Bell, Taylor J. and Burdanov, Artem and Cadieux, Charles and Charnay, Benjamin and Cloutier, Ryan and Cook, Neil J. and Correia, Alexandre C. M. and Dang, Lisa and Daylan, Tansu and Delrez, Laetitia and Edwards, Billy and Fauchez, Thomas J. and Flagg, Laura and Fraschetti, Federico and Haqq-Misra, Jacob and Huang, Ziyu and Iro, Nicolas and Jayawardhana, Ray and Jehin, Emmanuel and Jin, Meng and Kite, Edwin and Kitzmann, Daniel and Kral, Quentin and Lafrenière, David and Libert, Anne-Sophie and Liu, Beibei and Mohanty, Subhanjoy and Morris, Brett M. and Murray, Catriona A. and Piaulet, Caroline and Pozuelos, Francisco J. and Radica, Michael and Ranjan, Sukrit and Rathcke, Alexander and Roy, Pierre-Alexis and Schwieterman, Edward W. and Turner, Jake D. and Triaud, Amaury and Way, Michael J.},
	month = jul,
	year = {2024},
	pages = {810--818},
}

@article{moran_high_2023,
	title = {High {Tide} or {Riptide} on the {Cosmic} {Shoreline}? {A} {Water}-rich {Atmosphere} or {Stellar} {Contamination} for the {Warm} {Super}-{Earth} {GJ} 486b from {JWST} {Observations}},
	volume = {948},
	issn = {2041-8205, 2041-8213},
	shorttitle = {High {Tide} or {Riptide} on the {Cosmic} {Shoreline}?},
	url = {https://iopscience.iop.org/article/10.3847/2041-8213/accb9c},
	doi = {10.3847/2041-8213/accb9c},
	language = {en},
	number = {1},
	urldate = {2024-08-29},
	journal = {ApJL},
	author = {Moran, Sarah E. and Stevenson, Kevin B. and Sing, David K. and MacDonald, Ryan J. and Kirk, James and Lustig-Yaeger, Jacob and Peacock, Sarah and Mayorga, L. C. and Bennett, Katherine A. and López-Morales, Mercedes and May, E. M. and Rustamkulov, Zafar and Valenti, Jeff A. and Adams Redai, Jéa I. and Alam, Munazza K. and Batalha, Natasha E. and Fu, Guangwei and Gonzalez-Quiles, Junellie and Highland, Alicia N. and Kruse, Ethan and Lothringer, Joshua D. and Ortiz Ceballos, Kevin N. and Sotzen, Kristin S. and Wakeford, Hannah R.},
	month = may,
	year = {2023},
	pages = {L11},
}

@article{radica_promise_2025,
	title = {Promise and {Peril}: {Stellar} {Contamination} and {Strict} {Limits} on the {Atmosphere} {Composition} of {TRAPPIST}-1 c from {JWST} {NIRISS} {Transmission} {Spectra}},
	volume = {979},
	issn = {2041-8205, 2041-8213},
	shorttitle = {Promise and {Peril}},
	url = {https://iopscience.iop.org/article/10.3847/2041-8213/ada381},
	doi = {10.3847/2041-8213/ada381},
	language = {en},
	number = {1},
	urldate = {2025-01-13},
	journal = {ApJL},
	author = {Radica, Michael and Piaulet-Ghorayeb, Caroline and Taylor, Jake and Coulombe, Louis-Philippe and Benneke, Björn and Albert, Loic and Artigau, Étienne and Cowan, Nicolas B. and Doyon, René and Lafrenière, David and L’Heureux, Alexandrine and Lim, Olivia},
	month = jan,
	year = {2025},
	pages = {L5},
}

@article{bello-arufe_evidence_2025,
	title = {Evidence for a {Volcanic} {Atmosphere} on the {Sub}-{Earth} {L} 98-59 b},
	volume = {980},
	issn = {2041-8205, 2041-8213},
	url = {https://iopscience.iop.org/article/10.3847/2041-8213/adaf22},
	doi = {10.3847/2041-8213/adaf22},
	language = {en},
	number = {2},
	urldate = {2025-05-29},
	journal = {ApJL},
	author = {Bello-Arufe, Aaron and Damiano, Mario and Bennett, Katherine A. and Hu, Renyu and Welbanks, Luis and MacDonald, Ryan J. and Seligman, Darryl Z. and Sing, David K. and Tokadjian, Armen and Oza, Apurva V. and Yang, Jeehyun},
	month = feb,
	year = {2025},
	pages = {L26},
}

@article{weiner_mansfield_no_2024,
	title = {No {Thick} {Atmosphere} on the {Terrestrial} {Exoplanet} {Gl} 486b},
	volume = {975},
	issn = {2041-8205, 2041-8213},
	url = {https://iopscience.iop.org/article/10.3847/2041-8213/ad8161},
	doi = {10.3847/2041-8213/ad8161},
	language = {en},
	number = {1},
	urldate = {2025-05-29},
	journal = {ApJL},
	author = {Weiner Mansfield, Megan and Xue, Qiao and Zhang, Michael and Mahajan, Alexandra S. and Ih, Jegug and Koll, Daniel and Bean, Jacob L. and Coy, Brandon Park and Eastman, Jason D. and Kempton, Eliza M.-R. and Kite, Edwin S.},
	month = nov,
	year = {2024},
	pages = {L22},
}
\bibliographystyle{aasjournal}

\end{document}